\documentclass[english,12pt,a4paper]{article}
\usepackage{babel,amssymb,amsmath,graphicx,hyperref}

\title{Black Hole Shadow Observations with
Space-Ground Interferometers}
\author{E. V. Mikheeva, S. V. Repin, V. N. Lukash\\
\it{P.N. Lebedev Physical Institute of Russian Academy of Sciences}\\
\it{Profsoyuznaya 84/32, 117997, Moscow, Russia}}
\begin{document}

\maketitle

\begin{abstract}
We consider the black hole (BH) shadow images which can be restored by data processing and 
image recovery procedures in future space Very Large Baseline Interfe\-ro\-metry (VLBI) missions. 
For Kerr BHs with masses and coordinates of SgrA${}^*$, M87${}^*$ and M31${}^*$, 
illuminated by light source behind them, we consider three kinds of observation: the ground-based 
interferometer (similar to the Event Horizon Telescope), space-ground inter\-fe\-ro\-me\-ter 
with a satellite in geocentric orbit and space-ground interferometer with a satellite located in 
Lagrange point $L_2$. The significant difference between the images produced by the ground-based 
telescope alone and the one of the space VLBI with an added low-orbit satellite is caused by 
both the increased baseline and the improved of $(u,v)$ coverage. The near-Earth configuration of 
the radio interferometer for the observation of BH shadows is  the most preferable for the BH shadow 
observations among considered cases.  With further increasing the orbit radius up to the Lagrange point 
$L_2$ the density of the $(u,v)$ filling decreases and the results appear less reliable. 
Model images for all the cases are presented.
\end{abstract}

\section{Introduction}

Nowadays the observation of the black hole (BH) shadows is one of the hottest problems in
the astrophysics (see review \cite{Falcke_2017}-\cite{Nampalliwar_2018}). The pioneer 
observations of the BH shadow in the innermost area of M87 by the Event Horizon Telescope 
(EHT)\footnote{\texttt{http://eventhorizontelescope.org}} have just been made 
(\cite{EHT_collaboration_2019a}-\cite{EHT_collaboration_2019f}). 
The resulting images clearly demonstrate the exceptional capabilities of modern instruments. 
At the same time, progress in the study of black holes and their nearby surroundings requires 
the continuation of research, i.~e. the observation of BHs in other radio sources and at other (higher) 
frequencies.

At present, when the BH interferometry is developing very rapidly, the terms used to designate 
the BH images are: ``shadow'', ``silhouette'', and ``photon ring'' (the last term was first used 
in~\cite{Johnsonetal_2019}). Sometimes their meaning is shared, but often mixed. So, 
in~\cite{Dokuchaevetal_2019} the term ``silhouette''  is used, which is understood as the image of 
the BH event horizon, in \cite{Perlicketal_2015} the term ``shadow'' is used, which refers to 
the image of the so-called photon sphere ($r=3GM$ for a non-spinning BH). In English-language literature, 
terms ``shadow''and ``silhouette''are often confused. The authors follow this practice and name 
the shadow of BH its image for the specific model, without any detailing what the border of the shadow is. 
The image of the black hole depends both on its mass and rotation, and on the properties of the source 
that illuminates the black hole (disk, jet, bright spot, etc.). Some aspects of the building of the BH shadow 
in the General Relativity (GR) and extended theories of gravity have been discussed in \cite{Zakharov_2014}
(see also references therein), as well as in \cite{Zakharov_2018}-\cite{AlexeevProkopov_2020}.

However, regardless of the definition, the BH shadows have tiny angular sizes even for the closest BHs. 
This means that an VLBI technique should play a key role in shadow observations. Important parameters 
are also the magnitude of the baseline projection and the frequency at which the receiving equipment operates. 
The ground-based interferometers are limited by the Earth diameter and their baselines cannot exceed 12800 km. 
However, the great advantage of this construction is that we can relatively easy control it, repair, develop and 
collect the big data arrays. The ground-based VLBI array is realized in EHT array, 
which has been carried out the first observations of the BH shadow in the center of M87 at 230~GHz.

An angular resolution of interferometer can be improved by increasing the baseline or by drifting to 
higher frequency. So, it would be a more productive idea to build the array of space-based 
radiotelescopes, which could form an interferometer with huge baselines 
(\cite{Matveenko_1965}-\cite{An_2019}). A more detailed consideration of possible satellite orbits can be found in 
\cite{Fishetal_2019}-\cite{Palumnoetal_2019}. The angular resolution of that interferometer 
might thousand times exceed the one for the ground-based devices. But, being realized, the mission appears to be 
very expensive and its technical support can hardly be realized as easy as it is executed for the ground-based 
antennas. In addition to the technical and financial problems there are a lot of difficulties in the filling of $(u,v)$ plane. 
In ground-based observations it is possible to achieve a relatively homogeneous coverage of the $(u,v)$ plane 
because the overall configuration of these tracks is almost the same every 24 hours, whereas in space-ground 
observations, especially with large projections of the base, the coverage is expected to be rather heterogeneous. 
It leads to the additional difficulties in imaging procedure.

Nevertheless, the idea of space-ground interferometer is very tempting and now it is under heat consideration 
and discussion (see \cite{Fishetal_2019}-\cite{Palumnoetal_2019}). It can be realized in 
future space missions. One of them is the Millimetron space mission~\cite{Kardashev_2014} with cooled 
10-meter mirror operating in millimeter-~ and submillimeter bands, which is planned to be launched in late 2020s. 
The angular resolution of this instrument in the interferometric mode is assumed to be so high that we can clearly 
observe, in principle, the shadows of the BHs in many galaxies. However, there are reasons that worsen this ideal 
picture. For example, there are limitations on the sensitivity of the instrument, the radiation scattering by plasma 
inhomogeneities may occur, etc. In addition, all these effects depend drastically on the frequency. In the paper, 
we focus on the fundamental possibility of observing the shadow of a BH and neglect the nuances associated with 
the characteristics of specific observational instruments, as well as specific astronomical objects. The preliminary 
catalog of supermassive BHs can be found in \cite{Ivanov_2019}, where the observational possibilities of 
Millimetron mission were taken into account. The catalog is based on the extended catalog \cite{Mikheeva_2019}.

The interpretation of the interferometric observations of BH shadows requires the simulation of the expected image. 
This problem, in turn, requires a lot of effort to develop the radiation source models. As it is shown in many papers 
(\cite{Luminet_2018}, \cite{Luminet_1979}-\cite{Johannsen_2016}) the BH image depends significantly on 
the BH surrounding, i.~e. on the structure of the accretion disk, the dependence of its temperature on the radial coordinate, 
on the existence of relativistic plasma jet, etc. It is also necessary to take into account the radiation of the corona, 
the geometry of the magnetic field, the presence of synchrotron radiation, and more. Except that, the image depends 
also on the interstellar scattering processes (\cite{Zhu_2019},\cite{Johnson_2018}). In our simulation we do not take 
into account all these effects and use the simplest model of the shadow image, which depends only on the mass and spin 
of a BH and the source of photons - the bright plane behind the BH. Nevertheless, it allows us to reveal the characteristic 
features of the results of ground-based observations and the observations in space-ground interferometer. We do not 
consider an image of the real source, SgrA$^*$ (or M87$^*$, M31$^*$), but a model of a BH shadow with the same
angular size.

The recent observation of the BH shadow was provided by the ground-based interferometer. This means that 
we need to study the prospects for future research of BHs. The most evident way to do it is to join the experiences 
of EHT and RadioAstron mission \cite{Kardashev_2013} in future space experiments. In the paper we compare 
the images of the BH shadow which can be obtained by the interferometers with different configurations.

Main goal of the paper is to discuss the preferred satellite orbit, which should allow to obtain the shadow of a BH 
with high resolution quickly. Such a high-quality image is able to deliver the important information about the physical 
processes in the very vicinity of supermassive BHs (structure of the inner disk and base of a plasma jet), inhabiting the
innermost parts of massive galaxies. This also can be applied to test the General Relativity in strong gravitational fields. 
We do not discuss the technical problems of interferometry and leave their solution to VLBI specialists.

\section{Model of BH shadow}

\begin{figure}[t]
  \centerline{
  \includegraphics[width=8cm]{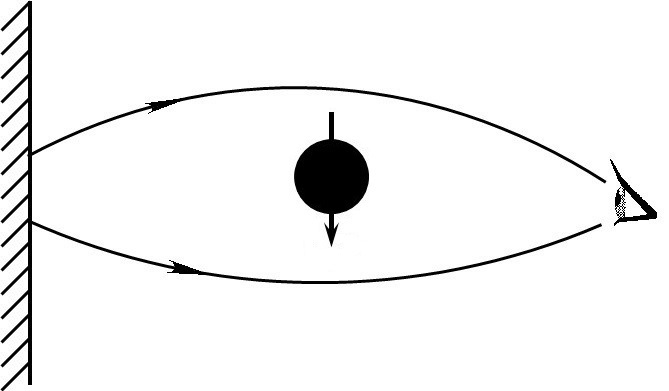}
              }
  \caption{
  The location of the emitting surface, the BH and the observer. 
                 The view line of the observer is perpendicular to both the screen plane 
                 and the spin axis of the BH.}                 
  \label{Figure_1}
\end{figure}

\begin{figure}[t]
  \centerline{
  \includegraphics[width=8cm]{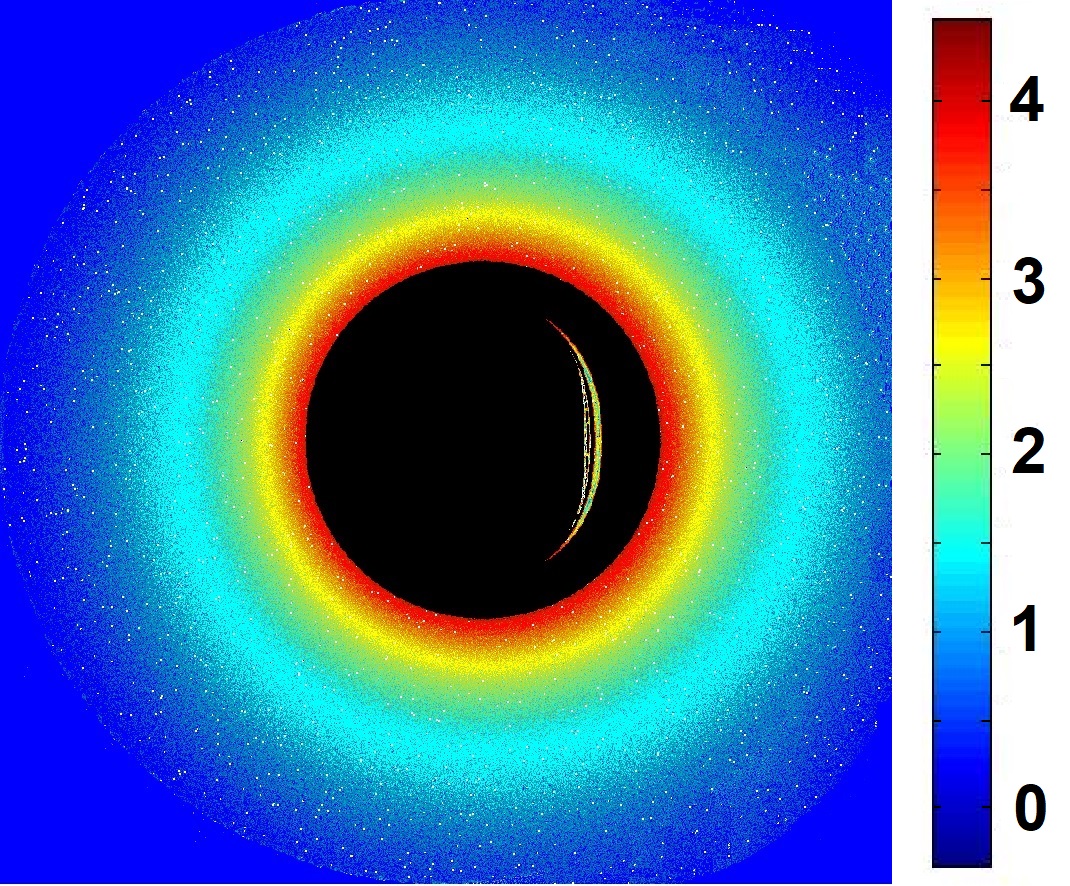}
              }
  \caption{
  The shadow of a Kerr black hole against the background of a flat luminous screen 
                 perpendicular to the line of sight of distant observer. The black hole spin axis is also 
                 perpendicular to the line of sight. The intensity is presented in logarithmic scale and 
                 normalized by the brightness of the background screen.
                 }                 
  \label{Figure_2}
\end{figure}

We consider the spinning BH and its shadow, or silhouette, with a simple geometry of photons source. 
We assume that a BH is described by the Kerr metric and its spin is close to maximal, $a = 0.9981$
(dimensionless parameter describing the ratio of the angular momentum of the BH to its mass).
The spin axis is perpendicular to the view line of the distant observer (see fig.~\ref{Figure_1}). 
Behind the BH and far away there is a bright flat screen, which emits the quanta uniformly to a hemisphere (in
solid angle $2\pi$). If the screen plane is perpendicular to the view line of the distant observer then the BH 
silhouette looks like the one shown in fig.~\ref{Figure_2}\footnote{If the BH was irradiated from a solid 
angle $4\pi$, the shadow image would look different: a wide dark ring between narrow photonic arcs and 
a contrasting edge (the red ring in the color system fig.~\ref{Figure_2}) would be light.}. 
The similar image might appear if a BH of the stellar mass orbits a red giant star and passes in front of the star.

To build the photon trajectories we solved the equations of motion under the General Relativity 
assumptions for each quantum. The system of six ordinary differential equations can be found 
in~\cite{Zakharov_1999, Repin_2018}. Ordinary differential equation solvers are included in many 
packages and freely distributed in Internet (\cite{Petzold_1983, 
ODE_solver_2018}\footnote{\texttt{https://www.mcs.anl.gov/petsc/}}). 
The simulated image counts the trajectories of approximately 10 million quanta.

This image has got a number of characteristic details: it is asymmetric, its brightness is inhomogeneous 
and, finally, it includes the thin annuli inside, formed by the quanta, which came to the observer after a few 
revolutions around the BH. The left hand side of the annuli is actually presents, but cannot be adequately 
displayed because the width of all the annuli is much smaller than the pixel diameter. On the right hand side 
the annuli can be seen separately and their total width is only 17-18 times less than the shadow diameter. 
Thus we consider the image of a BH shadow, which has enough small size details to elaborate and discuss 
the data processing technique. Certainly, such small and dim details will be highly likely blurred by the scattering
processes, but this fact will be considered later (see Conclusion).

We presume that the spin axis of the BH in the center of the Milky Way is perpendicular to the galactic plane 
and coincides with the axis of the Milky Way rotation. It means, that we know its orientation on the celestial 
sphere. As to M87 and M31, their spin axes and the spin axes of their BHs keep some uncertainty yet, so 
we accepted that their spins are oriented along the declination axis. Finally, we assume that our sources are 
time-independent.     

\section{Images of BH shadows}

To reconstruct the images we used the well-established {\sc{clean}} procedure (\cite{Hogbom_1974, 
CLEAN}). This algorithm is widely used in astrophysics and allows to extract the image from the Fourier 
coefficients on a finite number of $(u,v)$ plane points (which implies a smoothing procedure). 
Mathematically, we deal with an incorrect problem because a coverage of $(u,v)$ plane is incomplete. 
As it has been shown, for example, in \cite{EHT_collaboration_2019d}, the BH shadow images have some 
deviations between different image recovery methods and their different implementations. Nevertheless, 
as it has been demonstrated there, the morphology of images remains unchanged. In the paper, we also 
focus on the morphology of the image and do not consider the features that may be associated with 
the use of a specific procedure of image recovery or with the source model. The {\sc {clean}} method 
does give a general idea of the shadow image.

\subsection{Ground-based Interferometer}

\begin{figure}[t]
  \centerline{
  \includegraphics[width=55mm]{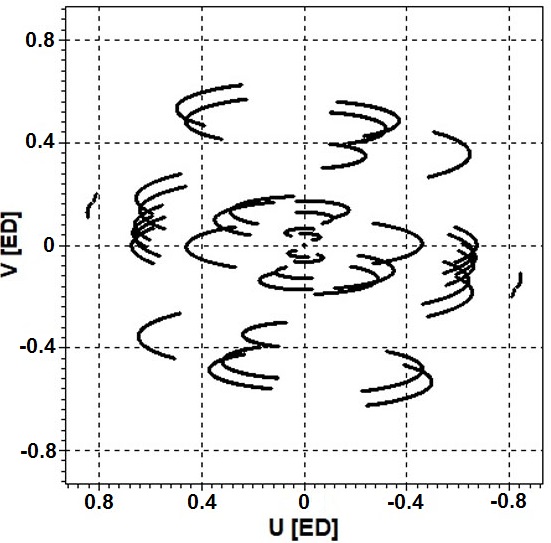}
  \includegraphics[width=55mm]{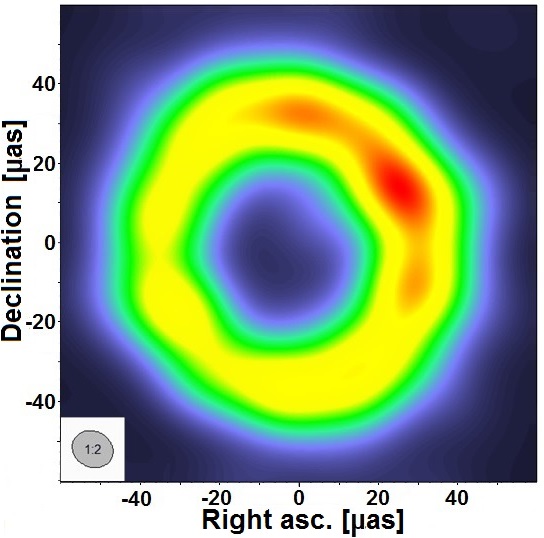}
              }
  \centerline{
  \includegraphics[width=55mm]{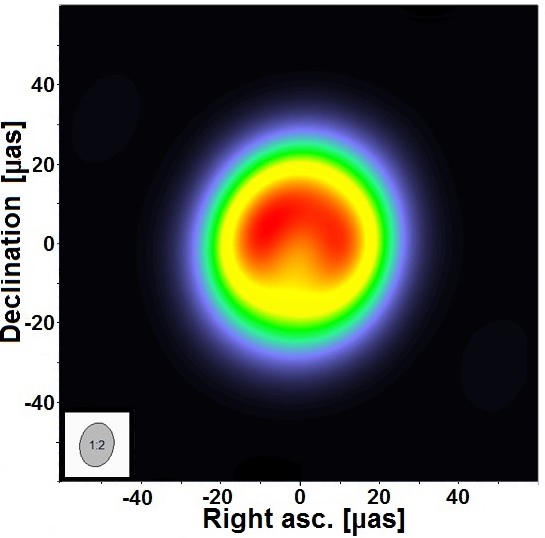}
  \includegraphics[width=55mm]{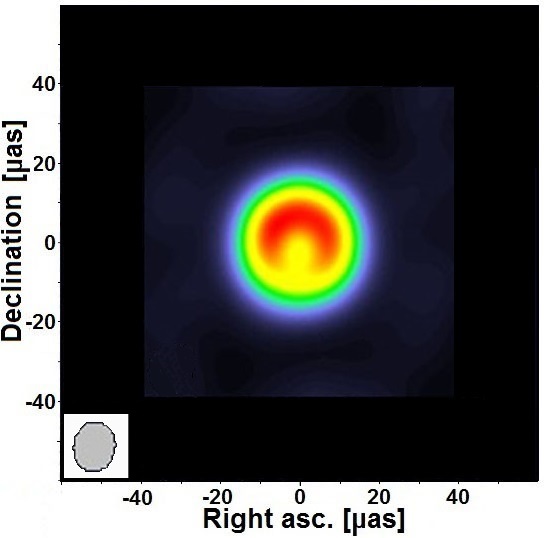}
              }
  \caption{
Images of a sample BH shadow for SgrA$^*$ (the top right panel), 
                M87$^*$ (bottom left) and M31$^*$ (bottom right) restored by data processing 
                for ground-based interferometric observations. The appropriate coverage of $(u,v)$ 
                plane is presented for SgrA$^*$. The images are shown in conditional colors.
                }                 
  \label{Figure_3}
\end{figure}

First, we consider the image, which can be reconstructed after the ground-based interferometric 
observations like the ones carried out by EHT. The daily coverage of $(u,v)$ plane for SgrA$^*$ 
shown in fig.~\ref{Figure_3} on the top left panel. The coverage of $(u,v)$ plane shown in 
the Figure is just a possible example and does not coincide with the real observational set of 
the objects by EHT. The maximal base projection here is about $0.8R_\odot$. 
For other objects the look of $(u,v)$ plane coverage may vary due to the different celestial
coordinates, but the general view remains approximately the same. As it follows from
the Figure the coverage is dense enough and looks relatively uniform. 

Three other panels in fig.~\ref{Figure_3} present the images of the BH shadow from 
fig.~\ref{Figure_2} observed at the frequency of 240~GHz and reconstructed then by {\sc{clean}}  
technique for coordinates and masses of SgrA$^*$, M87$^*$ and M31$^*$. As it follows from 
the Figure the resolution of the image is not high, but we can identify some details of the characteristic
image details, especially for SgrA$^*$. For example, at the top right corner there is a bright detail which 
can be interpreted as a narrow crescent, which can also be found in the original model image. 
On the restored images of other BH models (bottom panels of fig.~\ref{Figure_3}), the shadow is 
not visible. Asymmetry and heterogeneity of the image are an artifact of the data processing procedure.

\subsection{Low-orbit satellite Interferometer}

\begin{figure}[t]
  \centerline{
  \includegraphics[width=55mm]{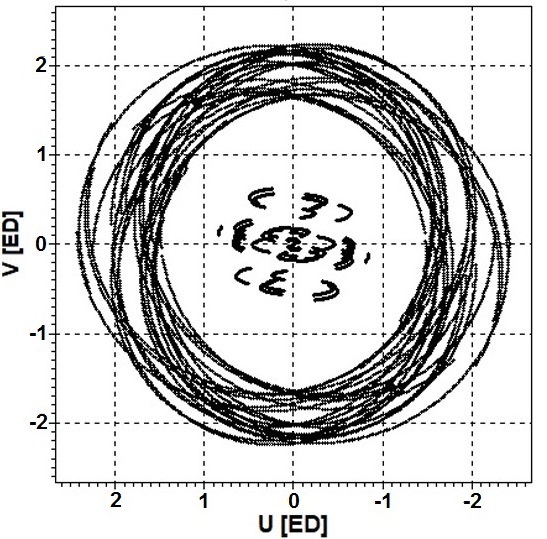}
  \includegraphics[width=55mm]{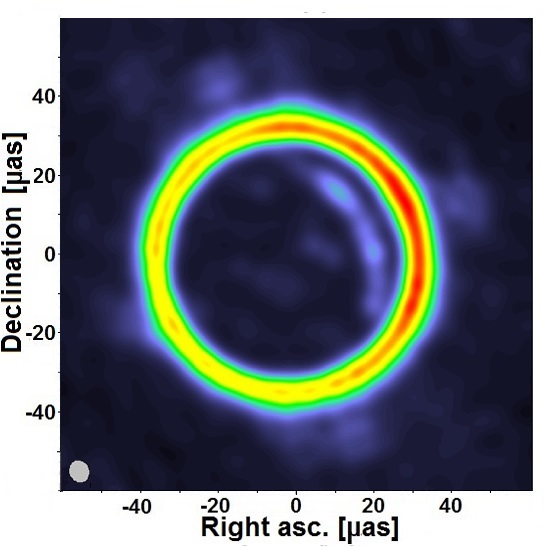}
              }
  \centerline{
  \includegraphics[width=55mm]{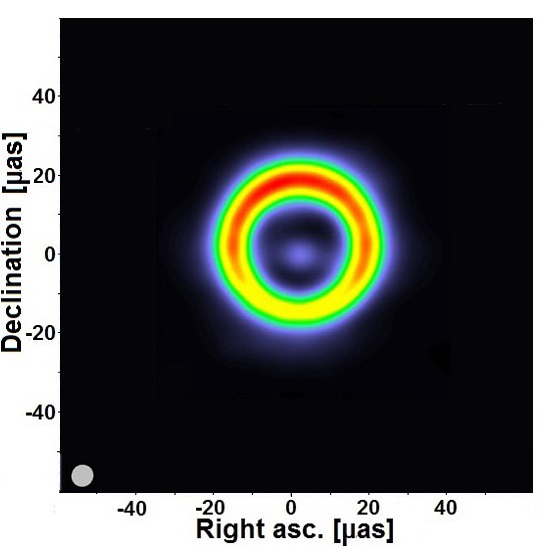}
  \includegraphics[width=55mm]{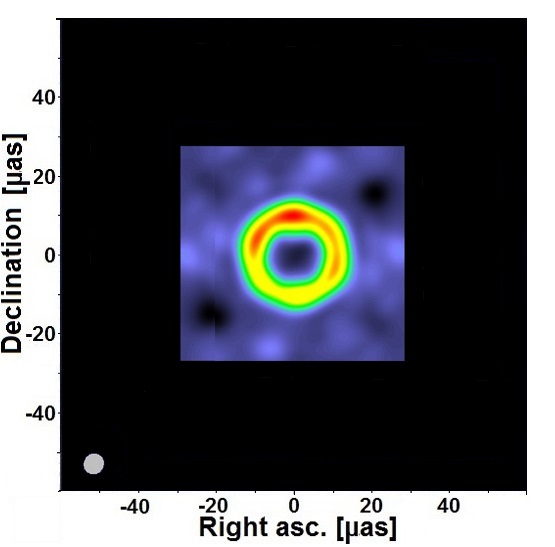}
              }
  \caption{
  Images of a sample BH shadow for SgrA$^*$ (the top right panel), 
                M87$^*$ (bottom left) and M31$^*$ (bottom right) restored by data processing 
                for space-ground interferometer with a satellite at low geocentric orbit. The appropriate 
                coverage of $(u,v)$ plane is presented for SgrA$^*$. The images are shown in 
                conditional colors.
                }                 
\label{Figure_4}
\end{figure}

The low-orbit interferometer implies the satellite at the orbits at 200-300 km from the Earth to 
the geostationary orbit. Their mean value radius is approximately $2\div3R_\oplus$. Each satellite 
in the array makes from 1 to 16 revolutions around the Earth per day, depending on the radius. 
In our simulation the radius of the orbit is close to $2R_\oplus$. The daily coverage of $(u,v)$ plane 
for SgrA$^*$ is shown on the top left panel in fig.~\ref{Figure_4}. The tracks of ground-based 
telescopes are also presented in fig.~\ref{Figure_4}, they are the same as those shown in 
fig.~\ref{Figure_3}, but in a reduced scale. As it has been above mentioned the general view of 
the coverage is approximately the same for M87$^*$ and M31$^*$.

The images of the objects, restored by the  {\sc{clean}} technique, are shown on three other panels in
fig.~\ref{Figure_4}. The top right panel presents the image of SgrA$^*$. This image contains a lot of 
additional details in comparison with fig.~\ref{Figure_3}. Thus, in the top right part of the inner dark 
area one can see the bright crescent that corresponds to the appropriate detail in fig.~\ref{Figure_2}. 
Moreover, we can even guess that this image is rotated at some angle compared to fig.~\ref{Figure_2} 
and measure immediately this angle. It gives us the important information on the momentum of the BH 
and the direction of its axis. It means that the angular resolution in this case is so high that after the image 
processing we can examine in detail the parameters of a BH and its environment. The bright circle has also 
the inhomogeneous intensity distribution and the position of its maximum is in agreement with the one of 
the inner crescent.

The images of M87$^*$ and M31$^*$ on the bottom panels in fig.~\ref{Figure_4}, have also 
additional characteristic features. It is clearly seen that the intensity of the bright ring is inhomogeneous
and this inhomogeneity carries the important information about the orientation of the BH spin axis. 
Higher angular resolution leads us to the fact that both the range of intensity in the images and 
the intensity gradients are much higher than in fig.~\ref{Figure_3}.

The comparison of the images in fig.~\ref{Figure_3} and~\ref{Figure_4} allows us to draw 
a conclusion that the low orbital space-ground interferometer, being realized, can get us much more 
information about the nature of the BHs than the observations of the ground-based interferometers only.

\subsection{High-orbit Satellite Interferometer}

\begin{figure}[tb]
  \centerline{
  \includegraphics[width=55mm]{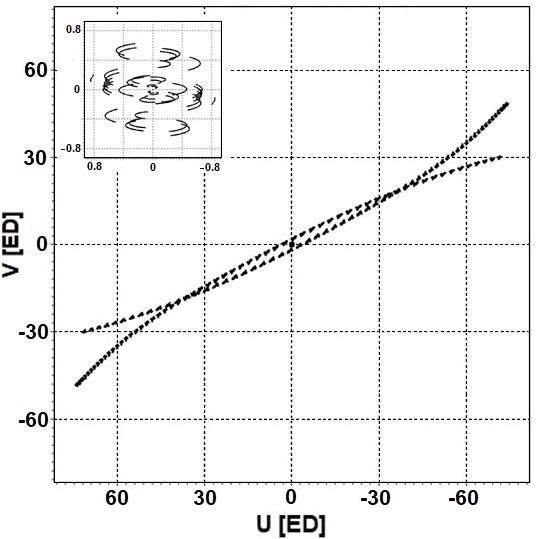}
  \includegraphics[width=55mm]{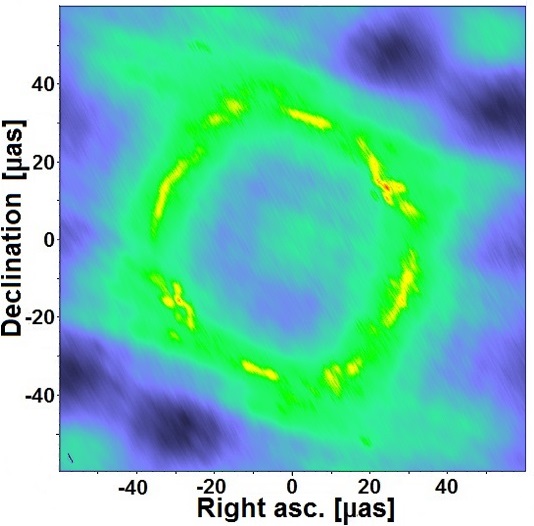}
              }
  \centerline{
  \includegraphics[width=55mm]{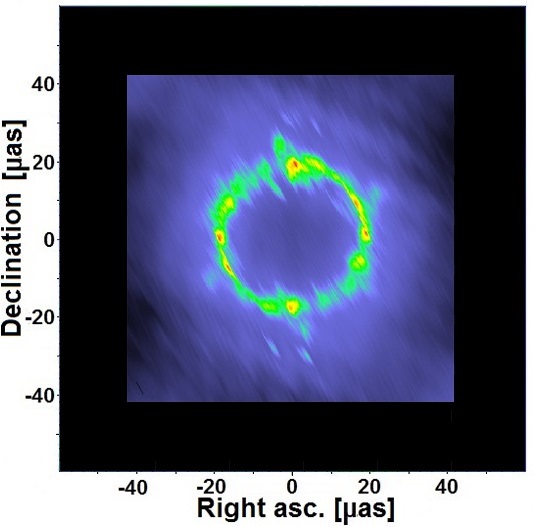}
  \includegraphics[width=55mm]{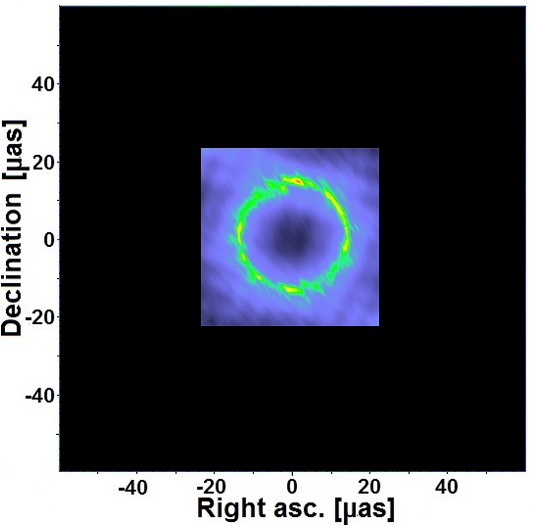}
              }
  \caption{
  Images of a sample BH shadow for SgrA$^*$ (the top right panel), 
                M87$^*$ (bottom left) and M31$^*$ (bottom right) restored by data processing space-ground 
                interferometer with a satellite located in Lagrange point $L_2$. The appropriate 
                coverage of $(u,v)$ plane is presented for SgrA$^*$. The images are shown in 
                conditional colors.
                }                 
  \label{Figure_5}
\end{figure}

The next step in increasing the baseline would be placing the satellite in the Lagrangian point $L_2$, 
that is the case of the planning space observatory Millimetron. After the satellite reaches this point 
it remains there during all the experiment time. The interferometer base can be really huge and 
its projection for some sources may exceed $100R_\oplus$. However, the coverage of $(u,v)$ 
plane becomes degenerated. Except that during the 5-year experiment one can hope for only 
a single observation of a specific object.
  
As before, the coverage of $(u,v)$ plane in this case for Srg A$^*$ is shown on the top left panel in
fig.~\ref{Figure_5}. The tracks of the ground-based antennas cannot be shown adequately because
of the fig.~\ref{Figure_5} scale. But they are really present in the very center of this Figure. They
are also presented in the inset. Notice that the low-orbit satellite is not included in the consideration here. 
The other panels in fig.~\ref{Figure_5} demonstrate the reconstructed images of SgrA$^*$, 
M87$^*$ and M31$^*$.

The angular resolution of this configuration is extremely high, but because of very poor coverage 
of the $(u,v)$ plane the result looks worse than in fig.~\ref{Figure_4}. The main reason is that this 
coverage of the $(u,v)$ plane does not permit to restore reliably the original image.
 
The bright photon ring can be easily distinguished in the image of SgrA$^*$, but despite of the very high 
angular resolution, the narrow inner crescent detail is completely lost, so any information about the axis 
orientation is lost too. The intensity distribution along the ring looks also uniform, in contrast to 
fig.~\ref{Figure_4}.

Analysis of the images of the considered BH shadow model with M87${}^*$ and M31${}^*$ 
coordinates reveals the same features. We can conclude only that a certain ring-like structure is definitely 
observed. The edge of the shadow in M87$^*$ also does not reveal any characteristic details that 
would allow to establish the direction of the angular momentum.

\section{Conclusions}       

We found that the interferometer, which includes both the ground-based telescopes and a low-orbit 
satellite has the advantage in comparison with other reviewed cases. The key role here plays an ability 
to fill the $(u,v)$ plane with high density during the relatively short time (a day or week). 
And a low-orbit interferometer is able to successfully solve this problem because the good 
coverage of the $(u,v)$ plane takes here less than a week. Moreover, if the orbit lays at about 300-400 
km above the Earth the revolution lasts 1.5~hour and after 5-7 cycles the coverage of the $(u,v)$ plane 
becomes dense enough. At that case the observation procedure may last only about 10 hours and, 
in principle, we can get the BH image with satisfactory quality and resolution twice a day. A satellite in 
the orbit with $R \sim 2R_\oplus$ has about three times longer orbital period, so a similar coverage of 
the $(u,v)$ plane will be achieved in about a day (see fig.~\ref{Figure_4}). The simultaneous use of two 
or three satellites can further reduce this time.

The ground-based arrays have an enough $(u,v)$ coverage, but the most objects cannot
be observed all day long because they are below the horizon during some period. At those
periods the coverage of $(u,v)$ plane is, certainly, stopped and we have the traces which
look like a half of ellipse (see fig.~\ref{Figure_3}). The great advantage of ground-based 
instruments is, obviously, their low cost.

The longer is a baseline of the interferometer the more difficult is the correlation process.
In particular, this is due to the difficulty in synchronizing of ground and on-board timers.
It means that the space-ground interferometer should have much more sophisticated equipment
than a ground instrument.

A case of a high-orbit satellite differs from others. Indeed, its angular resolution is
incredibly high. Theoretically, it is so high that we would be able to see the annuli inside
the shadow area in fig.~\ref{Figure_2} separately, i.~e. the resolution could be even greater than it is
reproduced in the Figure. However, we face here a problem with a poor $(u,v)$~plane coverage.
The fact is that the information obtained in the interferometric observation is not the image
of the object itself. It is a Fourier Transform of the true image at limited number points
of the $(u,v)$ plane. The reliable image reconstruction is possible only with a dense enough 
coverage of $(u,v)$ plane. One turn around the Earth of the satellite in Lagrangian point $L_2$
takes a full year. And even an increase in the duration of observations will not have a radical
effect on the coverage of the $(u,v)$ plane. A spacecraft moving in the orbit near point $L_2$
every year almost repeats its track on the $(u,v)$ plane. The displacement of the spacecraft
up and down from the ecliptic plane only increases slightly the coverage of the $(u,v)$ plane.

Another interesting idea is to place the radiotelescope at the Lunar pole \cite{An_2019}. 
The orbital period of the natural Earth satellite lasts about a month, so one third or a half of a year 
(a few turns around) is enough to cover the $(u,v)$ plane. In any case, this is much less than that 
is required for the satellite located at the Lagrange point $L_2$. However, for the implementation of 
such a colossal project there are still many technical problems to be solved.

At the present only two successful space VLBI missions have been implemented: the
VLBI \textit{Space Observatory Programme (VSOP)} \cite{Hirabayashi_1998} and 
\textit{RadioAstron} \cite{Kardashev_2013}. Both were equipped by the receivers in centimeter 
wave bands, so it means that we have not enough experience yet.

The long-awaited detections of the BH shadow by EHT is the first step to test the
General Relativity in strong gravitational fields. However, one has found that its quantitative
features are not sufficient to distinguish between BHs using different theories of gravity. It
is highlighting the fact that the great caution is needed when interpreting the BH images as the tests 
of General Relativity \cite{Mizuno_2018}. Since the BH shadow can be measured more precisely
by the space-ground interferometer than by the ground-based one, the space-ground VLBI
mission does allow to carry out the stronger tests of the General Relativity and the accretion
models.

\section*{Financial support}
This activity has been partly supported by Russian Foundation for Basic Research, grant
19-02-00199  and the RAS project KP 19-270 ``Questions of the origin and evolution of the Universe 
using methods of ground observations and space research". 

\section*{Acknowledgement}
The authors are grateful to Dr.~A.~Andreanov for his kind help in preparation of the
BH shadow images and numerous useful discussion. We are grateful to Dr.~V.N.~Strokov
Dr.~I.N.~Pashchenko for their help with the manuscript preparation. We also grateful to
Dr.~V.I.~Kostenko and Dr.~S.F.~Likhachev for useful discussions. One author (S.R.) is very
grateful to Dr.~O.N.~Sumenkova, Dr.~R.E.~Beresneva and Dr.~O.A.~Kosareva for the possibility of
fruitful working on this problem.


\end{document}